# Text/Graphics Separation for Business Card Images for Mobile Devices

**A. F. Mollah[+], S. Basu[*], M. Nasipuri[*], D. K. Basu[†]**

[+] *School of Mobile Computing and Communication, Jadavpur University, Kolkata, India*
*E-mail: afmollah@gmail.com*

[*] *Dept. of CSE, Jadavpur University, Kolkata, India*
*E-mail: subhadip@ieee.org, mnasipuri@cse.jdvu.ac.in*

[†] *AICTE Emeritus Fellow, Dept. of CSE, Jadavpur University, Kolkata, India*
*E-mail: dkbasu@cse.jdvu.ac.in*

## Abstract

Separation of the text regions from background texture and graphics is an important step of any optical character recognition sytem for the images containg both texts and graphics. In this paper, we have presented a novel text/graphics separation technique for business card images captured with a cell-phone camera. At first, the background is eliminated at a coarse level based on intensity variance. This makes the foreground components distinct from each other. Then the non-text components are removed using various characteristic features of text and graphics. Finally, the text regions are skew corrected and binarized for further processing. Experimenting with business card images of various resolutions, we have found an optimum performance of 98.54% with 0.75 MP images, that takes 0.17 seconds processing time and 1.1 MB peak memory on a moderately powerful computer (DualCore 1.73 GHz Processor, 1 GB RAM, 1 MB L2 Cache). The developed technique is computationally efficient and consumes low memory so as to be applicable on mobile devices.

## 1 Introduction

Pervasive availability of mobile phones with built-in cameras has drawn the attention of a number of researchers towards developing camera based applications for handheld mobile devices. Automatic Business Card Reader (BCR) is an example of such applications. Using this application, necessary contact information can be directly populated into the phonebook from the business cards. Although, such applications are commercially available in some mobile handsets, the accuracy is yet to be extended to be really useful in practice. Computational constraints of the mobile devices i.e. poor computing power, low working memory and no Floating Point Units (FPU) for floating point arithmetic operations add to the complexity of such systems.

It is observed that graphic backgrounds are commonly found in most business card images. In order to recognize the text information from the card, the text and background contents must be separated. The key challenges in that are the deformations in acquired images, diversity in nature of the business cards and most importantly the computational constraints of the mobile devices. The acquired image becomes skewed when the camera axis is not properly aligned with that of the object. When the planner objects are not parallel to the imaging plane, the captured images happen to be perspectively distorted. As both these adjustments are manually performed, perfection is quite difficult and so such distortion is very likely to happen, however small be it, for all camera captured business card images. In absence of sufficient light, one may opt for using camera flash. But, in that case, the centre of the card image becomes brightest and the intensity decays outward. The acquired images may be blurred if either the card or the camera is not perfectly static at the time of capturing or focussing is improper.

For practical applications, computational time and/or memory intensive algorithms can not be directly implemented on mobile devices in spite of their good performances. For this purpose, a computationally efficient yet powerful text separation method is designed in our work towards developing an efficient BCR for mobile devices.

## 2    A Brief Survey

Until recently, various text/graphic separation methods have been proposed and evaluated in regard to optical character recognition mostly for document images. Some have also been proposed for business card images captured with a built-in camera of a mobile device [1]-[3]. Few other text extraction methods are reported in [4]-[6]. DCT and Information Pixel Density have been used to analyze different regions of a business card image in [1]. Based on this analysis, the regions falling into the graphic components are identified and removed. In [2], a low resource consuming region extraction algorithm, which requires low computing resources such as processing speed, memory etc., has been proposed for mobile devices with the limitation that the user needs to manually select the area in which the analysis would be done. However, the success rate of this algorithm is yet to be improved.

Pilu et al. [3] in their work on light weight text image processing for handheld embedded cameras, proposed a text detection method that fails sometimes to remove the logo(s) of a card and the technique often mistakes parts of the oversized fonts as background and can not deal with reverse text i.e. light texts on dark background. In [4], text lines are extracted from Chinese business card images using document geometrical layout analysis method. Fisher's Discrimination Rate (FDR) based approach followed by various text selection rules is presented in place of mathematical morphology based operations in [5].

Yamaguchi et al. [6] have designed a digit extraction method for their work on telephone number identification and recognition from signboards. Roberts filter has been used to detect edges from the initial images. Then non-digit components are checked out according to some criteria based on various properties of digits. Hough transform has been used for skew and slant angle estimation.

While some of the above methods seem to be computationally expensive, some others need better accuracy. In this paper, we have presented a computationally efficient text/graphics separation method that works satisfactorily for camera captured business card images under the computing constraints of mobile devices.

## 3    The Present Work

The current work discussed in this paper may be subdivided into four key modules, viz. background elimination from the camera captured business card images, graphics separation, skew estimation for each text regions of the image and binarization of each skew corrected textual blocks. These are discussed in the following subsections.

### 3.1 Background Elimination

At first, the camera captured business card images are virtually split into small blocks. A block is part of either background or a foreground component. This classification is done on the basis of intensity variance within the block. The intensity variance is defined as the difference between the maximum ($G_{max}$) and the minimum ($G_{min}$) gray scale intensity within the block. It is observed that the intensity variance of a text block is considerably more than that of a background block. This has been the key idea behind the present approach. So, if the intensity variance of a block is less than an adaptive threshold ($T$), it is considered as part of the background. Otherwise, it is considered as part of a foreground component.

The threshold $T$ has two components, i.e. a fixed one ($T_{fixed}$) and a variable one ($T_{var}$) as shown in Eq. 1. $T_{fixed}$ is a constant subject to tuning. The formulation of $T_{var}$ is given in Eq. 2. It may be noted that $G_{min}$ must be greater than a heuristically chosen threshold $T_{min}$. Otherwise, the block is considered as part of a foreground object. This reveals the reality that even if the intensity variance of a block is less than $T$, it is not classified as background until the minimum intensity within the block exceeds $T_{min}$. This reduces the possibility of miss-classifying foreground blocks as background ones.

$$T = T_{fixed} + T_{var} \qquad (1)$$

$$T_{var} = [(G_{min}-T_{min}) - min(T_{fixed}, G_{min}-T_{min})] * 2 \qquad (2)$$

It is evident from Eq. 2 that the computation of $T$ is such that the more is the average intensity within the grid, the larger is the threshold. In other words, if the intensity band of the block falls towards the higher range of the overall intensity band, then $T$ becomes larger. Such formulation helps to efficiently eliminate the background blocks from the captured business card images. Also light backgrounds get easily eliminated in the said approach.

## 3.2 Graphics Separation

The standard 4-connected region growing algorithm [7] is applied to identify the distinct foreground Connected Components (CC) from background eliminated card images. A CC may be a picture, logo, texture, graphics, noise or a text region. In the current work, we focus to identify only the text regions using rule-based classification technique. The following features are used to classify a CC under consideration as a text region or not.

The height, width, width to height ratio (aspect ratio), gray pixel density, black pixel density, number of vertical and horizontal segments, and the number of cuts along the middle row of the CC are considered as features to decide upon the characteristic of each CCs. Different heuristically chosen adaptive (with respect to the size/resolution of the input image) thresholds are estimated for designing the rule-based classifier for text/graphics separation. Too small regions that are unlikely to become text regions and horizontal/vertical lines detected by checking their width, height and aspect ratio are considered as non-text components. Typically, a text region has a certain range of width to height ratio ($R_{w2h}$). So, we consider a CC as a potential text region if $R_{w2h}$ lies within the range ($R_{min}$, $R_{max}$). We assume that neither horizontal nor vertical lines can be drawn through a logo and it is larger than the largest possible character within the card. Thus, logos and other components satisfying the above specification get eliminated. Another important property of text regions is that the number of foreground pixels in a text region is significantly less than that of the background pixels. We consider a certain range of ratio of the foreground pixels to the background ($RA_{cc}$) given by ($RA_{min}$, $RA_{max}$) for the candidate text regions.

## 3.3 Skew Angle Estimation and Correction

Skewness in camera captured business card images may appear due to two primary reasons, firstly, misalignment of the handheld mobile camera with respect to the horizontal axis during image capturing, and secondly, due to perspective distortion within the captured image. In the later case, different text regions of any card image may be aligned at different skew angles with respect to the horizontal axis. To address this problem, in the present work, skew angles are estimated and subsequently corrected for each of the connected text regions.

To calculate the skew angle, we consider the bottom profile of a text region. Texts are surrounded by gray rectangular blocks found in Section 3.1. The profile contains the heights in terms of pixels from the bottom edge of the bounding rectangle formed by the text region to the first gray pixel found while moving upward. These heights are measured along the width of the region. However, if the extent of the gray shade along the column of a profile is too small, we discard as an invalid profile.

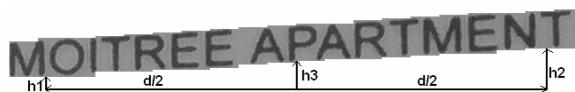

**Fig. 1: Skew angle computation for a typical text region**

The bottom profile of each text region is represented as an array of heights ($h[i]$s) from the bottom line, as shown in Fig. 1. The profile length is denoted as $N$. So, the mean ($\mu$) and the mean deviation ( ) are formulated as given in Eq. 3 and 4 respectively. Then we exclude the profiles that are not within the range (+ , - ). Among the rest of the profiles, the leftmost ($h1$), rightmost ($h2$) and middle ($h3$) heights are taken into consideration for skew angle computation. Thus, we find , and as the skew angles obtained from the pairs $h1$-$h2$, $h1$-$h3$ and $h3$-$h2$ respectively. The mean of them is considered as the computed skew angle of the text region. Skew correction is shown in Fig. 2 for some sample text regions extracted from different business card images.

$$\mu = \frac{1}{N} \sum_{i=0}^{N-1} h[i] \qquad (3)$$

$$\tau = \frac{1}{N} \sum_{i=0}^{N-1} |\mu - h[i]| \qquad (4)$$

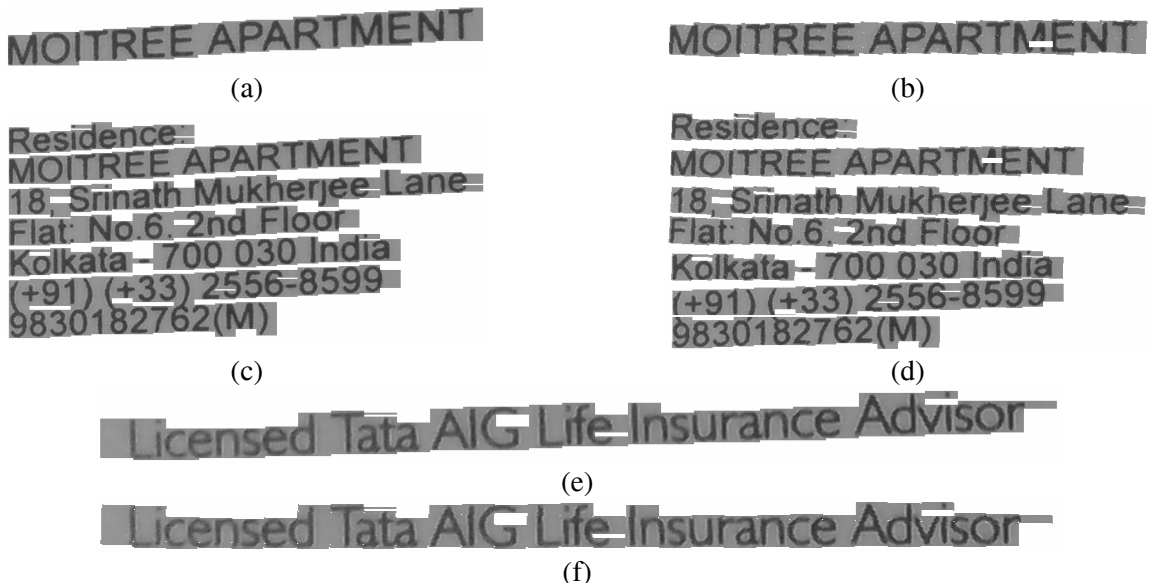

(a)

(b)

(c)

(d)

(e)

(f)

**Fig. 2: Sample text regions are shown before and after skew correction**
  (a) Single line connected component
  (b) Skew corrected look of (a)
  (c) A text region containing multiple lines
  (d) Deskewed text region for (c)
  (e) Large single text line
  (f) View of (e) after skew correction

## 3.4 Binarization

After skew correction of the text regions the images are binarized as follows. If the intensity of any pixel within a CC is found to be less than the mean of the maximum and minimum intensities of that CC, it is considered as a foreground pixel. Otherwise, we first check the 8 neighbors of the said pixel and if any 5 or more neighbors are already considered as or likely to be foreground; then we consider the pixel as a foreground one. All the remaining pixels are considered as background. The advantage of this approach of binarization is that the disconnected foreground pixels of a character are likely to become connected due to neighborhood consideration. Fig. 3 shows typical business card images with their graphics eliminated binarized snapshots.

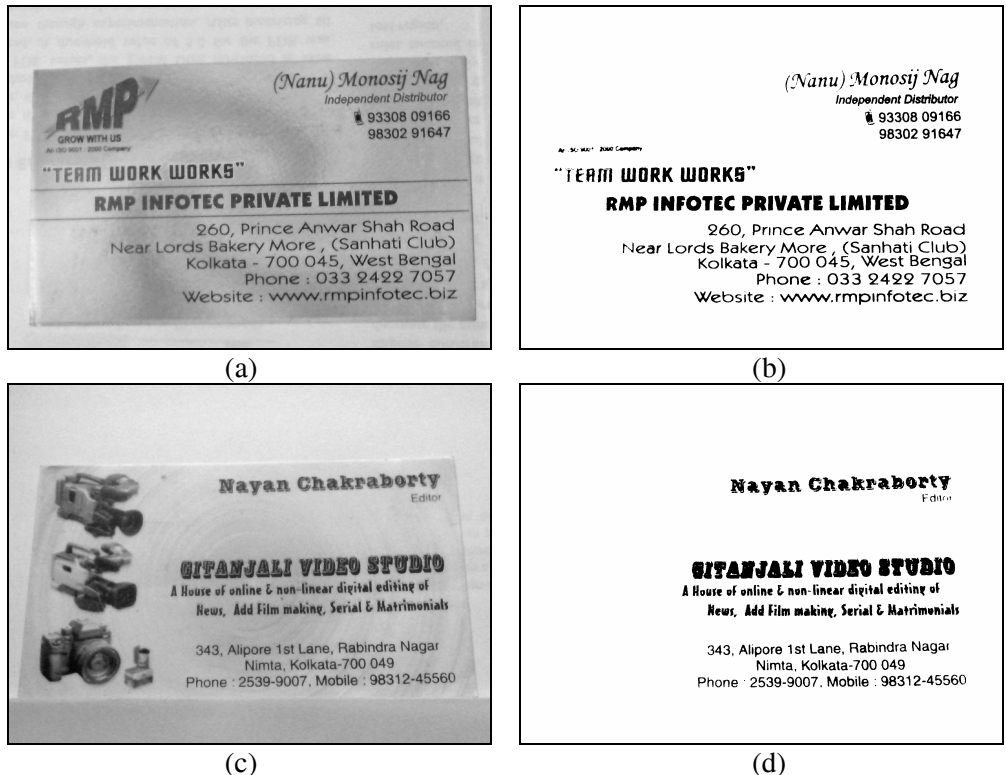

(a)  (b)

(c)  (d)

**Fig. 3: Sample card images and their graphics eliminated snapshots**
    (a)  A business card image with textual logo and multi-colored background
    (b)  Extracted texts from the card image (a)
    (c)  A business card image containing multiple pictures in the background
    (d)  Graphics eliminated view of card image (c)

## 4    Experimental Results and Discussion

To evaluate the performance of the present technique for text/graphics separation, experiments have been carried out on a dataset of 100 business card images of various types, acquired with a cell-phone camera (Sony Ericsson K810i). The dataset contains both simple and complex backgrounds including complex backgrounds and logos. Some cards contain multiple logos and some logos are combination of text and graphics. Most of the images are skewed, perspectively distorted and degraded. Sample images from business card dataset are shown in Fig. 3(a), 3(c), 4(a), 4(c) and 7(a).

### 4.1 Text/Graphics Separation Accuracy

To quantify the text/graphics separation accuracy, we have designed the following method. A graphic component here refers to all non-text regions including background texture and noises. All the remaining connected regions are identified as texts. While estimating the classification accuracy, four different situations may arise. Table 1 shows all such possible cases.

| CC/Region | Identified as | Justification |
|-----------|---------------|---------------|
| Background | Background | BB (True) |
| Background | Text | BT (False) |
| Text | Background | TB (False) |
| Text | Text | TT (True) |

**Table 1: Justification of classification rules for the CCs**

The text extraction accuracy of the current technique is defined in Eq. 5 where BB, TT are considered as true classifications and BT, TB are misclassified ones.

$$\text{Accuracy} = \frac{\text{No. of true classifications}}{\text{Total no. of CCs}} \qquad (5)$$

Using the above accuracy quantization method, we have got a maximum success rate of 98.93% for 3 Mega pixel images with $T_{fixed} = 20$, $T_{min} = 100$, $R_{min} = 1.2$, $R_{max} = 32$, $RA_{min} = 5$ and $RA_{max} = 90$. Table 2 shows the accuracy rates when experimented with other resolutions on a moderately powerful desktop (DualCore 1.73 GHz Processor, 1 GB RAM, 1 MB L2 Cache). Fig. 3-4 show that the technique works satisfactorily.

| Resolution (width x height) | Mean accuracy (%) |
|---|---|
| 640x480 (0.3 MP) | 97.80 |
| 800x600 (0.45 MP) | 97.86 |
| **1024x768 (0.75 MP)** | **98.54** |
| 1182x886 (1 MP) | 98.00 |
| 1672x1254 (2 MP) | 98.45 |
| 2048x1536 (3 MP) | 98.93 |

**Table 2: Classification accuracy of business card image with various resolutions**

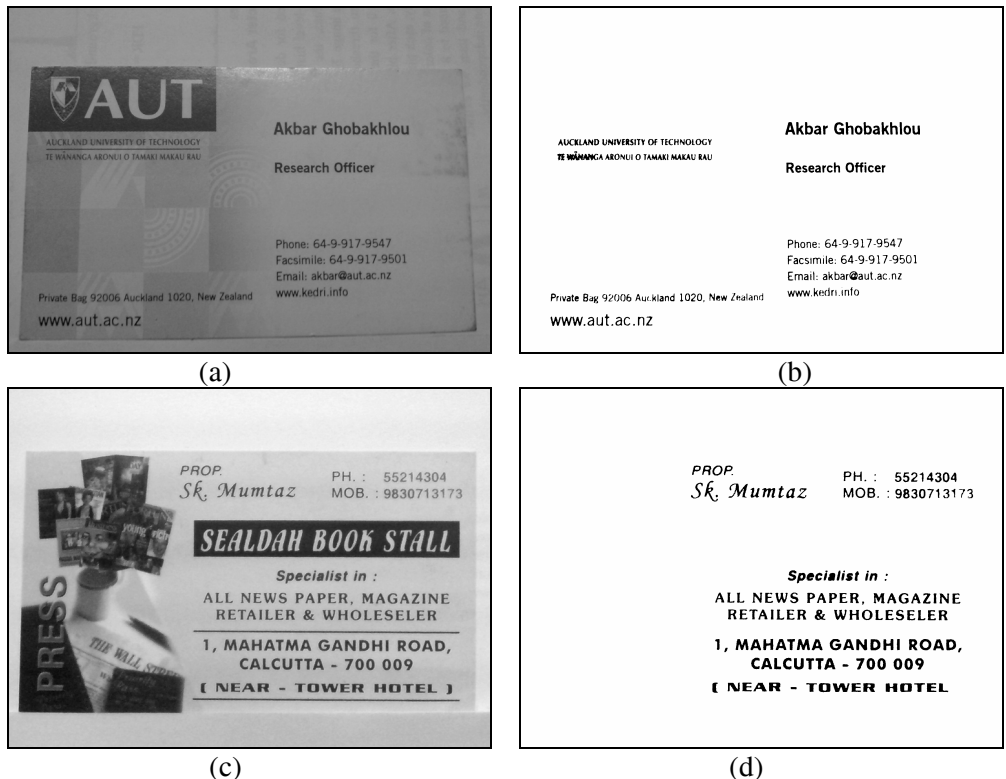

(a)

(b)

(c)

(d)

**Fig. 4: Successful text/graphics separation for business card images**
    (a) A dark business card image with unevenly patterned background
    (b) Background and non-text components eliminated view of card image (a)
    (c) A bright business card containing picture and a light text on dark background
    (d) Dark texts on light background are extracted from card (c)

## 4.2 Computational Requirements

The applicability of the presented technique on mobile devices is evaluated by its computational requirements. As our aim is to deploy the proposed method into mobile devices, we want to develop a light-weight Business Card Reader (BCR) system beforehand and then to embed into the devices since the computational challenges are known and one can easily have a decent idea about the timing and memory requirements for the proposed technique.

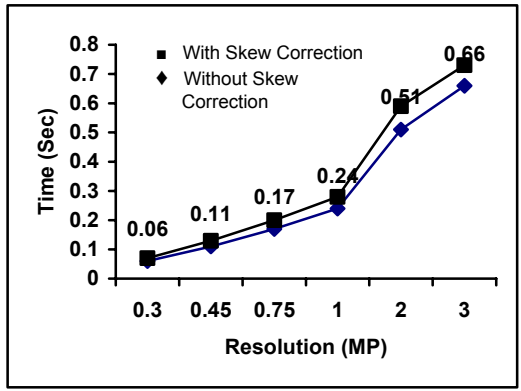

**Fig. 5: Time Computation with Various Image Resolutions**

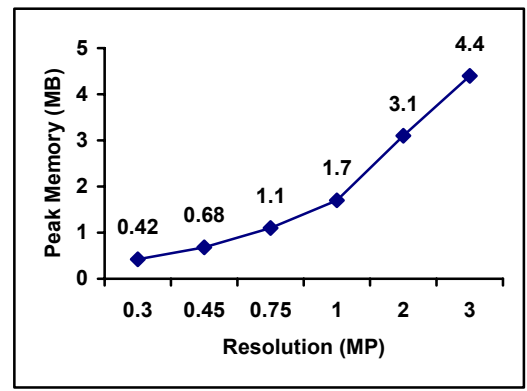

**Fig. 6: Memory Consumption with Various Image Resolutions**

An observation [8] reveals that the majority of the processing time of an Optical Character Recognition (OCR) engine, embedded into a mobile device, is consumed in preprocessing including skew correction and binarization. Although, we have shown the computational time of the presented method with respect to a desktop, the total time required to run the developed method on mobile devices will be tolerable. Fig. 5 and 6 show the timing and memory requirements of the present technique with various resolutions respectively. It may be observed from Fig. 5 that time requirements for skew correction technique, an essential component for the overall efficiency of the BCR, is minimal in comparison to the time required for text/graphics separation and binarization modules combined.

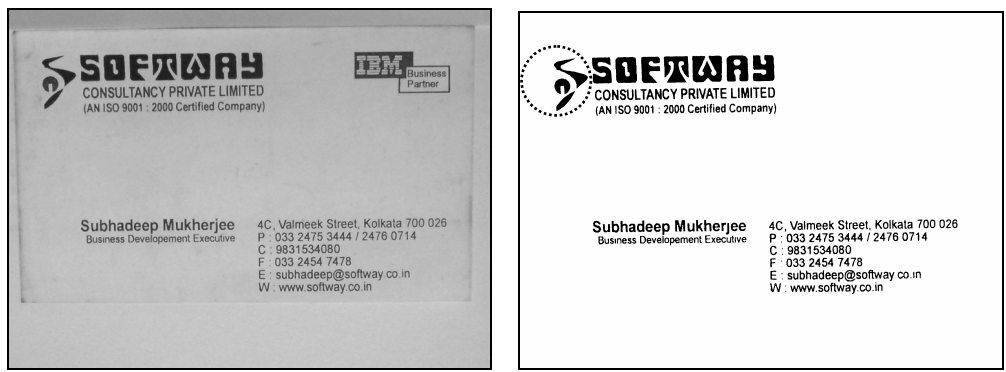

(a)             (b)

**Fig. 7: Sample business card images with partial misclassification**
(a) An input image with graphics close a text region
(b) Failed to remove graphics components encircled with dotted lines

## 4.3 Discussion

Although, one can observe that the developed method works well as shown in Fig. 3, Fig. 4 and Section 4.1, it too has certain limitations. Sometimes, the dot of 'i' or 'j' gets removed during CC analysis. When text and image/logo are very close to each other, they together often form a single CC and get wrongly classified as background or text as shown in Fig 7. Also, textual graphics (e.g. logo of different companies) get easily classified as potential text regions. The current technique is not applicable for white texts on a dark background as shown in Fig. 4(c-d). These areas may however be considered as our future directions of research in this field. However, text regions are hardly classified as graphics ones. So, by the present technique, we have the least possibility to loose any textual information in the BCR system. On the other hand, the graphics elements that are classified as texts may be subsequently eliminated in further steps like segmentation, recognition and post-processing. These steps are yet to be incorporated in our continuing work.

## 5    Conclusion

In the current work, we have developed a fairly accurate text/graphics separation methodology for camera captured business card images. The present work also implements a fast skew correction technique for the misaligned text regions and subsequent binarization of the gray scale background/graphics suppresses images. Although, the computational requirements have been shown with respect to a PC, the same may be acceptable for mobile architectures (200 MHz - DualCore 333 MHz ARM Processors, 22-128 MB RAM) as well. Observation reveals that with the increase in image resolution, the computation time and memory requirements increase proportionately. It is evident from Fig. 5 that the computationally efficient skew correction algorithm contributes to an average of 15% rise in the overall computation time. Although, the maximum accuracy is obtained with 3 mega pixel resolution, it involves comparatively high memory requirement and 0.66 seconds of processing time. It is evident from the findings that the optimum performance is achieved at 1024x768 (0.75 MP) pixels resolution with a reasonable accuracy of 98.54% and significantly low (in comparison to 3 MP) processing time of 0.17 seconds and memory requirement of 1.1 MB. In brief, the current work may be viewed as a significant step towards development of effective segmentation/recognition technique for designing a complete and efficient BCR system for mobile devices.

## Acknowledgment

Authors are thankful to the *Center for Microprocessor Application for Training Education and Research* (CMATER) and project on *Storage Retrieval - Understanding of Video for Multimedia* (SRUVM) of the Department of Computer Science and Engineering, Jadavpur University for providing infrastructural support for the research work. We are also thankful to the *School of Mobile Computing and Communication* (SMCC) for proving the research fellowship to the first author.